\documentclass{article}

\usepackage{amssymb}
\usepackage{graphicx}
\usepackage[linewidth=1pt]{mdframed}
\usepackage{lipsum}

\usepackage{url}

\begin{document}


\title{Computational Properties of Fiction Writing and Collaborative Work}

%
%
\author{Joseph Reddington (1), Fionn Murtagh (2) and  Douglas Cowie (3) \\
(1) Dept. of Computer Science, (3) Dept. of English, \\
Royal Holloway, University of London, UK\\
(2) School of Computer Sci. \& Informatics, \\
De Montfort University, Leicester, UK \\
{\tt j.reddington, douglas.cowie@rhul.ac.uk, fmurtagh@acm.org}}

%
%

\maketitle

\newcommand{\naive}{na\"\i ve }
\newcommand{\naively}{na\"\i vely }

\begin{abstract}
From the earliest days of computing, there have been tools to help shape narrative.  Spell-checking, word counts, and readability analysis,  give today's novelists tools that Dickens, Austen, and Shakespeare could only have dreamt of.  However,  such tools have focused on the word, or phrase levels. In the last decade, research focus has shifted to support for collaborative editing of documents.
This work considers more sophisticated attempts to visualise the semantics, pace and rhythm within a narrative through data mining.  We describe real life applications in two related domains. 
\end{abstract}


\section{Introduction}
\label{intro}
This work considers sophisticated attempts to visualise pace and rhythm within a narrative. 
The key insight of these techniques is not to replace a qualitative evaluation (the reading of the text) with a quantitative assessment, but, by means of a rigorous deterministic process, to extract relationships from input data and display them for interpretation.  In essence, one qualitative evaluation (of the text) is augmented with another (of an image); however, the qualitative evaluation of the image has the advantage that it is not only vastly faster, but also independent of both language and reader familiarity. 

Fiction writing is a competitive industry, and supports several sub-sectors in the form of writing classes, manuscript consultants, and networking events. Writers face challenges in getting feedback on their work, particularly in terms of rhythm and pace.  Not only is quality subjective, the process is extremely time-consuming for the reader.  Moreover, if the writer is to iterate through drafts of their work, then the feedback of any given reader becomes less and less useful as the reader becomes familiar with the text. There are also situational difficulties, such as if the writer simply doesn't accept aspects of the criticism as valid. 

A \naive tool might  split a narrative into chapters and then plot a chart showing how a measure like the Flesch reading index~\cite{flesch1948new} changed between chapters. Such a chart would have limited general use; however, if a chapter had a significantly different  index it would be sensible to conclude that the chapter was considerably different in style to the surrounding chapters and that the writer should be aware of this.   A key point here is that the writer certainly shouldn't be expected to change the narrative simply because one chapter is somewhat unusual by some measure. There are many possible  sensible reasons for the anomaly, but it is our position that it is to the writer's advantage that they are aware of both the result and the tool, so they can reason about why the result occurred.  If the writer has purposely caused the effect to further the narrative, then such a result would be a validation, otherwise, if the writer has accidentally caused the effect then they can consider the worth of the effect and potentially take steps to adjust or remove it.

This work uses a framework for narrative analysis proposed in~\cite{Murtagh2009302} and applies such techniques to two example domains, with a view to evaluating the system to see if it can provide insights of value in literary research.  One domain is in the traditional agent/consultant model, whereas the other is a group process, situated much closer to writing for TV or film scripts.

Of course, our comparisons are not an adequate or complete way of assessing individual style; they are nonetheless an element that can be employed usefully for our specific purpose.

This paper first details work in related areas and places the techniques examined here in an insightful and innovative context.   The following sections describe the operational use domains.  Visualisations of the narrative mapping are described.  The analysis of these mappings is accompanied by examples and notes on how the use was suited, or not, to particular aspects of each domain.  

\section{Previous Work}
\label{related}

Previous work relating narrative and computer science tends to focus on creation -- for example, designing systems that produce emergent narrative~\cite{callaway2002narrative,kriegel2008emergent,louchart2008purposeful} or by modelling an existing narrative as a sequence of actions with pre and postconditions~\cite{Porteous:2010}.   There are also many instances where media outlets have announced computer systems that can pick the next bestselling book, script, or music~\cite{nextbigthing}. The failure of these systems to live up to the hype has led people to be naturally cautious about any analysis system in the creative domains. 

The techniques examined in this paper were first used in~\cite{Murtagh2009302,Murtagh2010253} to distinguish the style and structure of film and TV scripts.  Murtagh et al.\ focused on capturing the semantics of the data and the plausibility of taking text as a practical and useful expression of underlying story. This work can be characterised as providing a platform to construct visual representations of the semantics encoded in the data. 

There is an overlap with the area of {\em sentiment analysis}, which analyses user-generated content: often by determining if the author of a  blog comment or tweet is in favour of, or against, a product.  Although visualisations have been constructed this way~\cite{Chen_abstractvisual,gamon2005pulse,kakkonen-galickakkonen:2011:DigHum}, such approaches are based, thus far, in examining a small set of sentiment-bearing words, and they consider the source text as a single block, rather than a set of discrete scenes comprising a narrative arc.

\section{Methodology and Evaluation Domain} 
This section presents details of domains of deployment.  Later sections will evaluate how different information mapping methods are used to enhance the workflow of each. Our evaluation is based on our observations and testimonials that were provided. 

A number of interviews were conducted with experts in the publishing industry that made it clear that there was a large degree of resistance to what the industry might see as ``replacement by robots''. The two mapping through visualisation techniques we evaluate here are of interest because they require a level of interpretation from the user, and so may be much more acceptable to the industry.

The use of these techniques was evaluated in two domains, which were selected to represent the extremes of creative writing. The Writer's Desk is a consultancy offering a very traditional feedback mechanism to authors, whereas Project TooManyCooks  models the deadline-driven high intensity creativity found in group writing for TV, film, or magazines.

\subsection{Project TooManyCooks}
Project TooManyCooks (TMC) (described briefly in~\cite{murtagh2011}) is a creative writing project that runs camps of 8 to 10 student writers who collaboratively create a novel (depending on the age of the students this is normally in the 30,000 to 65,000 word range) over a period of five days. It has two core goals: to  increase the contact time and feedback between students interested in fiction writing; and to give students experience of the  lifecycle of the novel from inception to printing.  Example outputs include~\cite{TMCdelivery,TMCdeception,TMCplaying,TMCroad}.  In this domain, users  were particularly interested in using the analysis techniques to quickly alert them to sections that in some sense didn't follow the overall voice of the rest of the novel.  The project was also interested in mapping and visualisation of overall plot arcs: allowing them to reorder sections in such a way that particular scenes do not overshadow each other within the narrative.

\subsection{The Writer's Desk} 
The Writer's Desk (TWD) is a commercial entity specialising in the review of manuscripts for authors~\cite{twd}.  TWD's role is in giving professional feedback to authors over the style and structure of their work.  This study spent six months providing narrative analysis for a selection of the submissions they received.  The analysis reports were either used internally for developing TWD feedback or passed on to authors as an appendix\footnote{An example of a report prepared for TWD can be found at http://www.cs.rhul.ac.uk/home/joseph/hosted/angel.pdf}.   TWD and their writers  were particularly interested in seeing the chapter-to-chapter flow and, as an extension of this, how an author's work sits as a whole.  As a commercial enterprise, TWD was also interested in identifying target markets -- and in grooming submissions to  hit an area of particular interest to the public more precisely. 

\section{Visualisations}

This work reports experiences using two mappings to express the narrative arc.  Firstly, quite general frequency of occurrence data is determined for word usage in context.  Based on all interrelationships between words and text segments, a mapping is obtained that is Euclidean and hence easly visualised as a map-like representation.  From that, and aided greatly by the Euclidean map -- most often, of full inherent dimensionality and hence not suffering any loss of information -- a tree or hierarchical visualisation is obtained.  A further innovative development is to have such a hierarchy respect a given ordering of the input text related to narrative development or chronology.

Each input text is automatically divided into a number of segments, with chapter headings being used to delimit segments\footnote{If there are no chapters in the text, but sections are divided by some distinct typographical convention, then section boundaries may be used instead.}. Given these segments and a list of unique words in the input text, a cross-tabulation is constructed which gives the count of the occurrences of a given word in a given segment.    From a machine-learning perspective our data was semi-structured, in that it is organised into discrete chapters or segments.  

One can use correspondence analysis to extract from a cross-tabulation some level of structure from the text in the form of an embedding in Euclidean space. Details of the construction are available in~\cite{murtagh2011,Murtagh2009302}. We refer to the extracted structure as mapping the semantics of the text, because each word is a weighted average of text segments, and each text segment is a weighted average of the words it contains. Both the tree visualisations to be presented use Euclidean space embedding as a starting point.  For each visualisation, a description is given with examples and then a detailed analysis is reported on of the advantages and also limitations of the visualisation in each domain.

\subsection{Unordered or Geometric}
\begin{figure}
\begin{center}
\includegraphics[width=6cm]{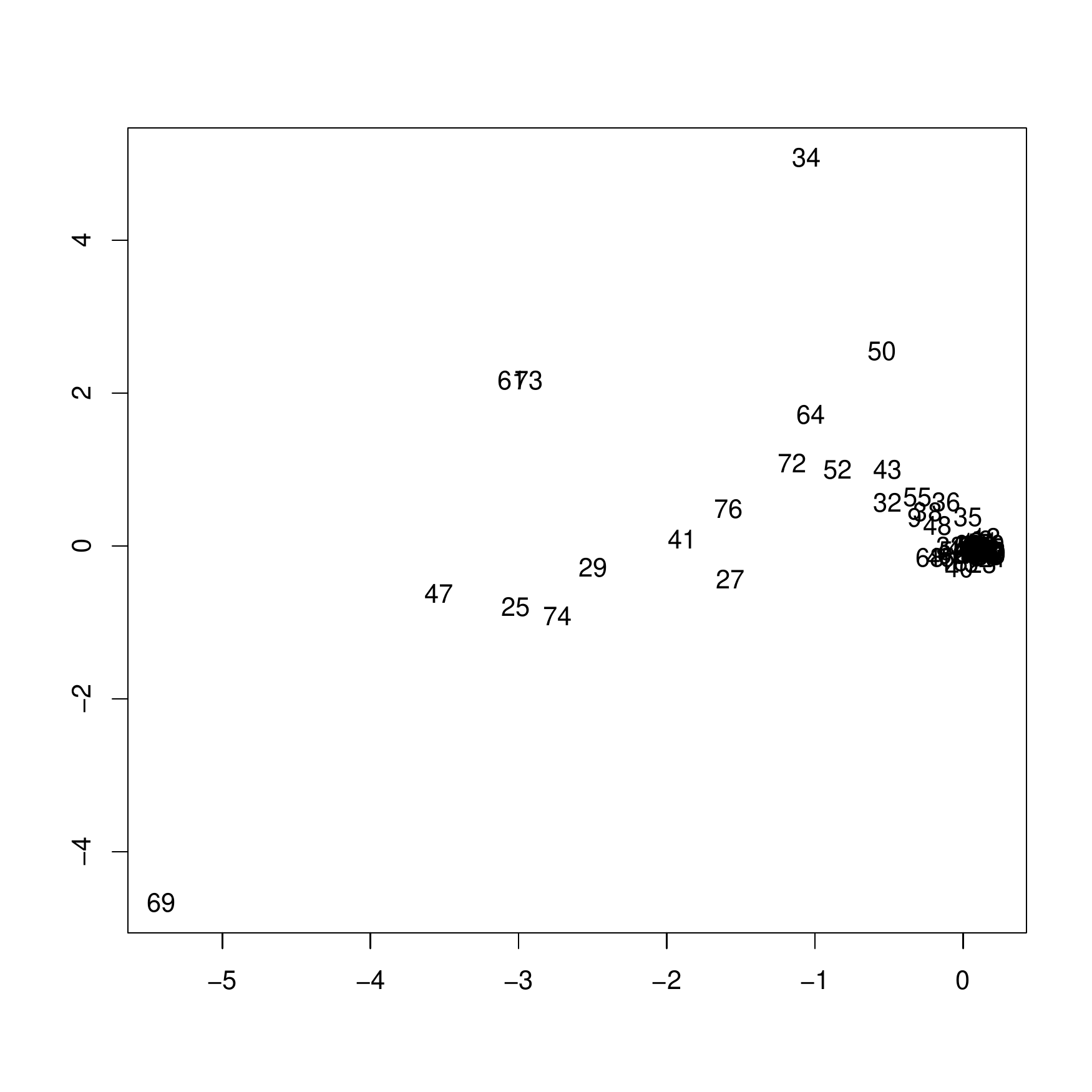}
\caption{{\em Owen Noone and the Marauder} as a geometric, 
best planar projection, visualisation.}
\label{default}
\end{center}
\end{figure}

The relationships in the data given by the set of all frequency of occurrence (including 0 = no presence) values can be projected into two dimensions to show the relative position of each chapter (text segment) in the projection. Figure~\ref{default} shows such a projection from {\em Owen Noone and the Marauder}~\cite{Cowie200502}, with each segment of the text represented by a point on the projection.  Since the process used the relative word counts as its starting point, two segments in the novel will appear closer to each other in the projection if they have similar relative word frequencies. It is our position that when an author writes a segment in a distinctly different style or tone (examples might be moving to a different tense or a sudden change in the tension in the storyline) then these word frequencies will change significantly and be visible on the projection for interpretation.  

For example, Figure~\ref{default} shows a tight core grouping over to the right hand side of the projection, with a number of outliers. Subjectively one might say that this grouping represents the ``voice'' of the author or novel -- and it may be considered worthwhile to investigate the nature of those segments that did not fit in with this voice. If one examines segment 69, 
which is the most extreme outlier, it can been seen that it is written as a fictional extract from the newspaper {\em USA Today}, as opposed to the majority of the novel, which is written with a more conventional third-person narrator.  The author very much intended to give this segment a different ``voice''. In this particular work, the majority of the other relative outliers are similar plot devices in the form of radio announcements, magazine articles and so on.   Of course, the software makes no judgement here. It simply displays the information for an expert evaluation. 

The example of {\em Owen Noone} is a static study of a published novel after a rigorous proofing and editing process. We shall shortly show how TWD used the visualisation to examine a snapshot of styles to position a novel in the market, while TMC used the visualisation to track the progress of construction over time.  

\subsubsection{TooManyCooks}

One of the core goals within the TooManyCooks process was to give the appearance of having one single author with a clear style and ``voice''.  The group originally relied on the ``wikipedia effect'' -- that is that if enough different authors proofread and rewrite the same section repeatedly, then differences in style become invisible to the causal reader.   However the 2-dimensional projection allowed users to visualise the style and see which sections might benefit from a stylistic rewrite.

It is tempting to assume that this ``core style'' was simply the average of the styles of the writers. In fact, this was the working model used in  TwoManyCooks -- this visualisation was introduced in the proofreading stages as a way of applying a consistent style across the novel. During the latter two days of a TooManyCooks project, the current draft of the novel is repeatedly printed out, proofread, and has changes made to it (generally on the order of three iterations per day). In early iterations, group coordinators would identify outliers -- evaluate each of them to see if the outlier was an intentional outlier and, if not, paired it with another segment that was in some sense opposing the first.  The group members who wrote the first drafts of each of these were instructed to copy-edit each other's draft with the intention that the stylistic differences would cancel out.   One could imagine a similar process pairing writers and sub-editors on a magazine or a newspaper. In later iterations, this becomes much more a process of identifying unintentional outliers and focusing the stronger writers on those chapters for rewriting, while other writers polished more minor corrections in those chapters that hadn't shown as outliers.

\begin{figure}
\begin{center}
\includegraphics[width=6cm]{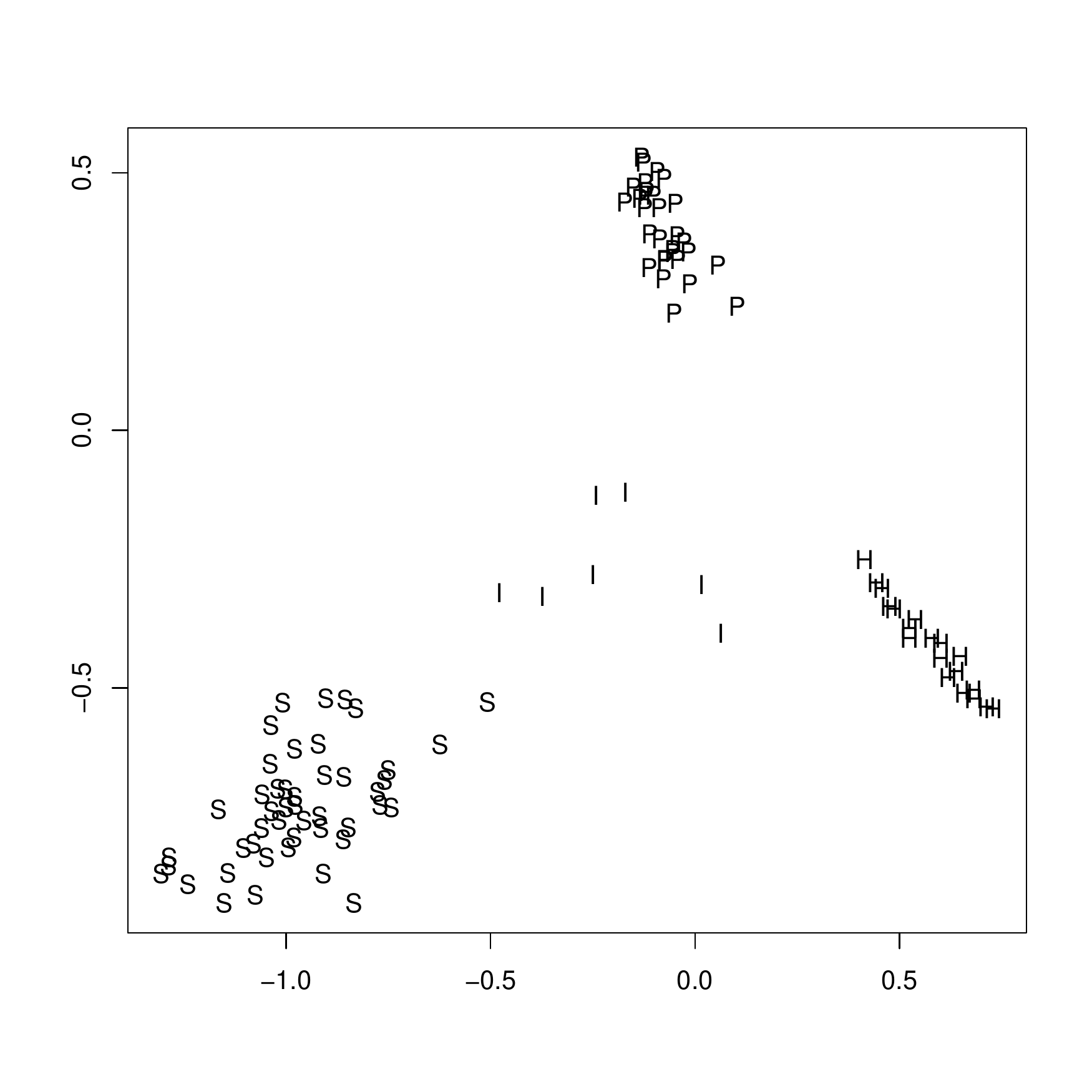}
\caption{Comparing author short stories (I) with their collaborative novel (S) and  commercial examples (H and P)}
\label{strodesScatter}
\end{center}
\end{figure}

Later work provided more grist for the mill of our thinking. A recent TooManyCooks group was selected from students who had won a short story competition. Figure~\ref{strodesScatter} shows a projection in which the short stories are compared with both the novel that the writers produced, and (for context) the popular novels {\em Harry Potter and the Half Blood Prince}~\cite{halfblood}, represented by H, and {\em Pride and Prejudice} \cite{Austen201205}, represented by P.   The short stories, represented by the I symbols, unexpectedly do not surround the novel that the authors later collaborated on (represented by S symbols).  This suggests that in fact the core clustering is more a result of the group of writers improving the consistency of their prose with regard to an intended style, rather than being shackled to a literary fingerprint.  Note also that the clustering of the TooManyCooks novel is much less tight than either of the two popular authors, which is probably to be expected from a small group of 6th-form students\footnote{16-17 years old.} writing over a five day period.  

The major use of the unordered visualisation for the TooManyCooks project was in identifying sections of unusual style and being able to evaluate each for its role in the story. Being able to highlight those aspects of the story that did not have the same ``voice'' as the main narrative allowed the writers to streamline the feedback process and present to readers a more consistent narrative.  

\subsubsection{The Writer's Desk (TWD)} 
An attraction of the projections for TWD was the ability to quickly compare with other artists within the same genre.  A regular complaint of publishers and agents is that they are sent manuscripts for genres in which they do not specialise and end up rejecting the vast majority of these out of hand.  At the fine-grained level, editors have regularly commented that an author does not necessarily write in the style that they believe they do and, more crucially, they do not necessarily aim at a market segment that they are best suited for.   By using the projection visualisation to compare a target manuscript with a selection of commercial novels one can compare explicitly. 

\begin{figure}[h]
\begin{center}
\includegraphics[width=6cm]{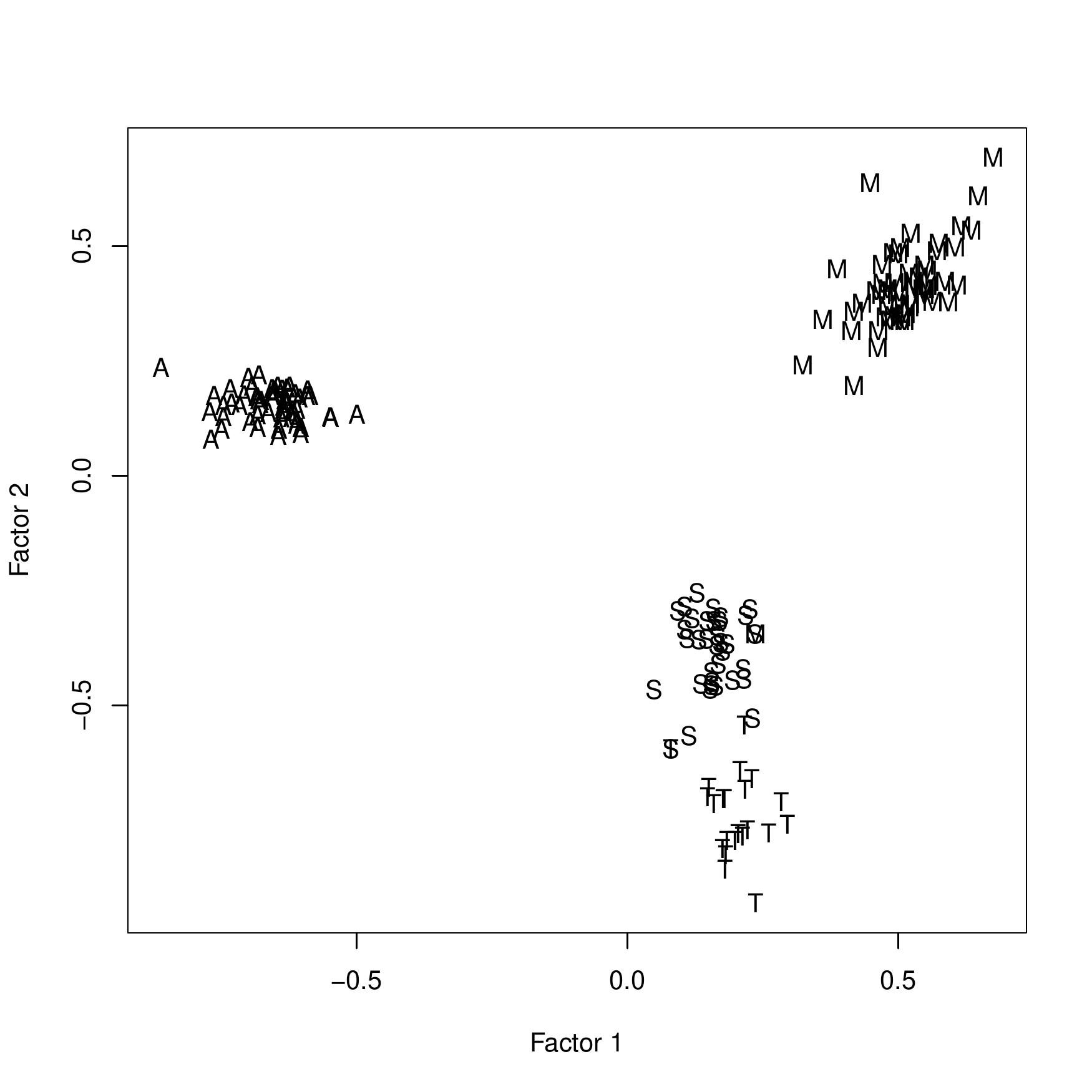}
\caption{Comparing a novel (T) with its commercial competitors.}
\label{track}
\end{center}
\end{figure}

For example, TWD had a commission to  examine a particular target novel that was aimed at the style of romance novel exemplified by Danellie Steel.  Figure~\ref{track} shows the the target novel text (T), compared with several other novels. These are: {\em Kaleidoscope}, by Danellie Steele (S); {\em Emma}, 
by Jane Austen (A); and {\em Eclipse}, by  Stephanie Meyer (M)~\cite{Steel198811,emma,Meyer200708}.  This allowed  TWD to evaluate, to their own satisfaction, if  the style and word choice in this instance was closer to the Steele-style romance than either the classic or teen styles of the other examples. Furthermore, the overall consistency of the text is similar to what would be expected from a published novel.  There is, of course, a psychological component to some of this feedback. Some authors react viscerally to the idea of this sort of analysis, fearing that the approach reduces creativity, while some react very favourably, having more faith in their own interpretation of the visualisations than they necessarily have in their agents or editors (who they might see as sparing them hard truths).

The ability to highlight anomalous sections was also of great interest within the TWD domain as it provided a useful metric for working one-on-one with authors, and to invite them to interpret the results in relation to their work. This allowed the conversation to be more about the guiding of the author and not about a difference in personal tastes between people.  

Feedback from the company was universally positive, particularly in the area of how comfortable they were in interpreting the visualisation for themselves, and in helping with the more commercial aspects of the business.

\subsection{Ordered or Hierarchical}
Although the planar presentations are useful, they do not address the fact that the narrative is consumed linearly, and so they reflect only those differences that we are referring to as style or mood between any given successive pair of segments.  To gain more insight into the actual structure of the narrative, a visualisation is used that respects the sequentiality of the segments. This section evaluates this hierarchical arrangement of the information, again starting from quite generic text/word association data with relatively minimal pre-processing..  

The hierarchical clustering algorithm used here is detailed in~\cite{murtagh1985multidimensional}, and was used as a device to deconstruct the film {\em Casablanca} in~\cite{Murtagh2009302}. Briefly, the algorithm repeatedly merges the least dissimilar pair of adjacent scenes to form a tree-like structure that shows how segments of a narrative cluster together. 

This sequential ordering allows the viewer to notice how, although a chapter or set of chapters may fit within the overall `style' of a novel, they may not necessarily match with their immediate neighbours. Once again, there can be outliers, and a human can decide if an outlier is there to intentionally shape the narrative or not. 

\begin{figure}[h]
\includegraphics[angle=0,width=130mm]{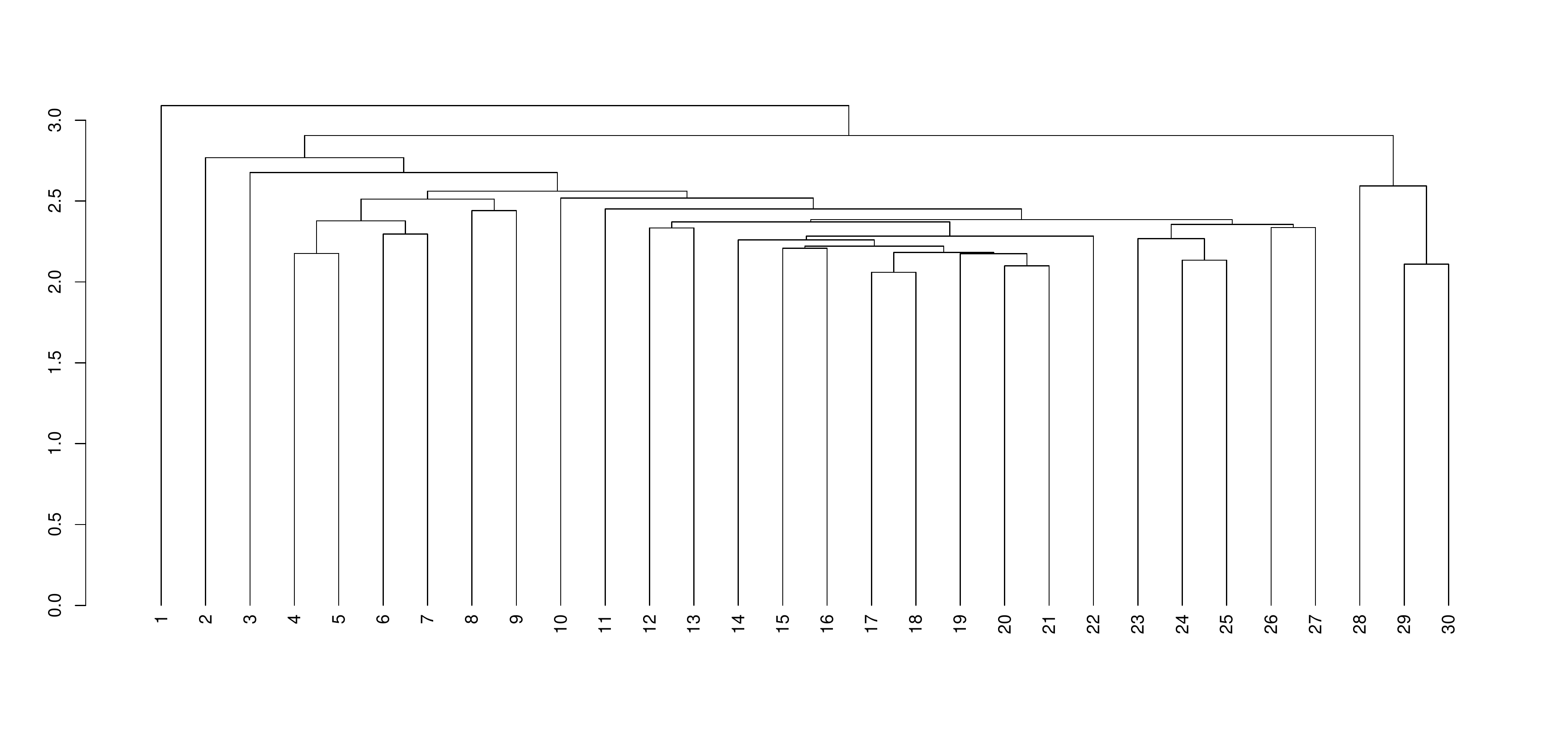}
\caption{Ordered visualisation of chapters in {\em Harry Potter and the 
Half-Blood Prince}.}
\label{PotterDen}
\end{figure}

For example, Figure~\ref{PotterDen} shows the ordered visualisation of {\em Harry Potter and the Half-Blood Prince} by J.K. Rowling~\cite{halfblood}, in which each segment is a chapter in the novel. Viewing the structure, one can see that  the cluster comprising only the first chapter is rated as being remarkably dissimilar to the cluster containing all other chapters.  The opening chapter of the novel is a conversation between the Prime Minister of the UK, and the Minister for Magic; the chapter is used mainly for setting up the narrative and the mood, and neither character features significantly in the remainder of the text.  A subjective reading of the novel may support that the first chapter was separate structurally from the text.  Although the comparative deconstructing of such works to a much lower level of detail is a fascinating subject in its own right, it is outside the scope of this work.   In particular, our two target domains focus much more heavily on the use of this ordered visualisation for examining novels as works-in-progress. For more information on this clustering see, e.g.~\cite{murtagh1985multidimensional,murtagh2005correspondence}.

\begin{figure}[h]
\includegraphics[angle=0,width=130mm]{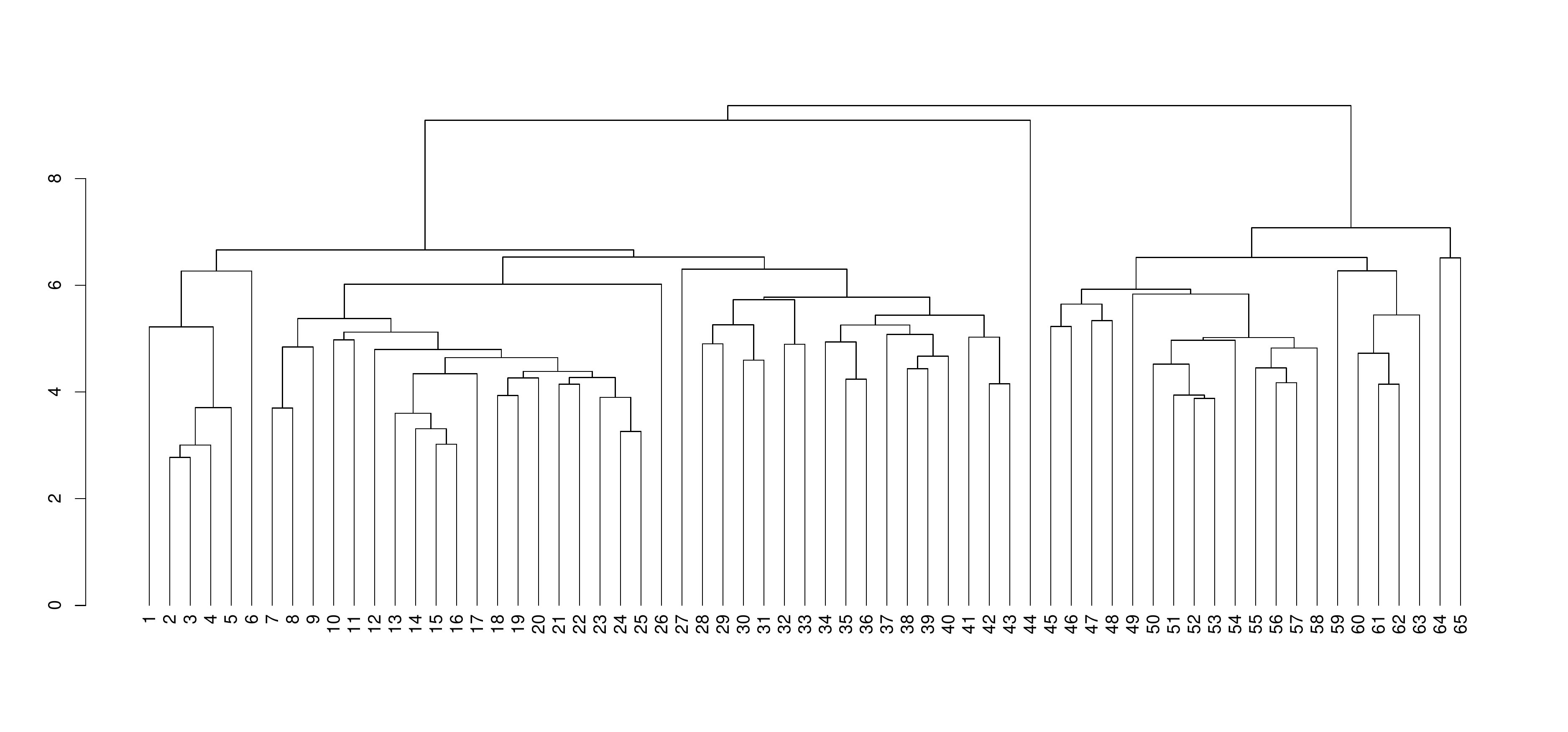}
\caption{Chapter structure of  {\em The Shadow Hours}, first snapshot 
from drafts.}
\label{mden2}
\end{figure}

\subsubsection{Project TooManyCooks}

Figure~\ref{mden2} shows a dendrogram using an early draft of {\em The Shadow Hours}, which was the test novel for the TooManyCooks Project.  The major anomalous section in Figure~\ref{mden2} (chosen by eye) is 44, followed by  26, 27, and 6.   Section 44 happens to be the smallest section in the narrative in that draft -- and the only one that hadn't been expanded from a skeletal outline  into a draft section so it required attention.  Sections 26 and 27 were character development of one of the minor characters; they had been drafted by one team member and had not been reviewed yet by other team members.

\begin{figure}
\includegraphics[angle=0,width=130mm]{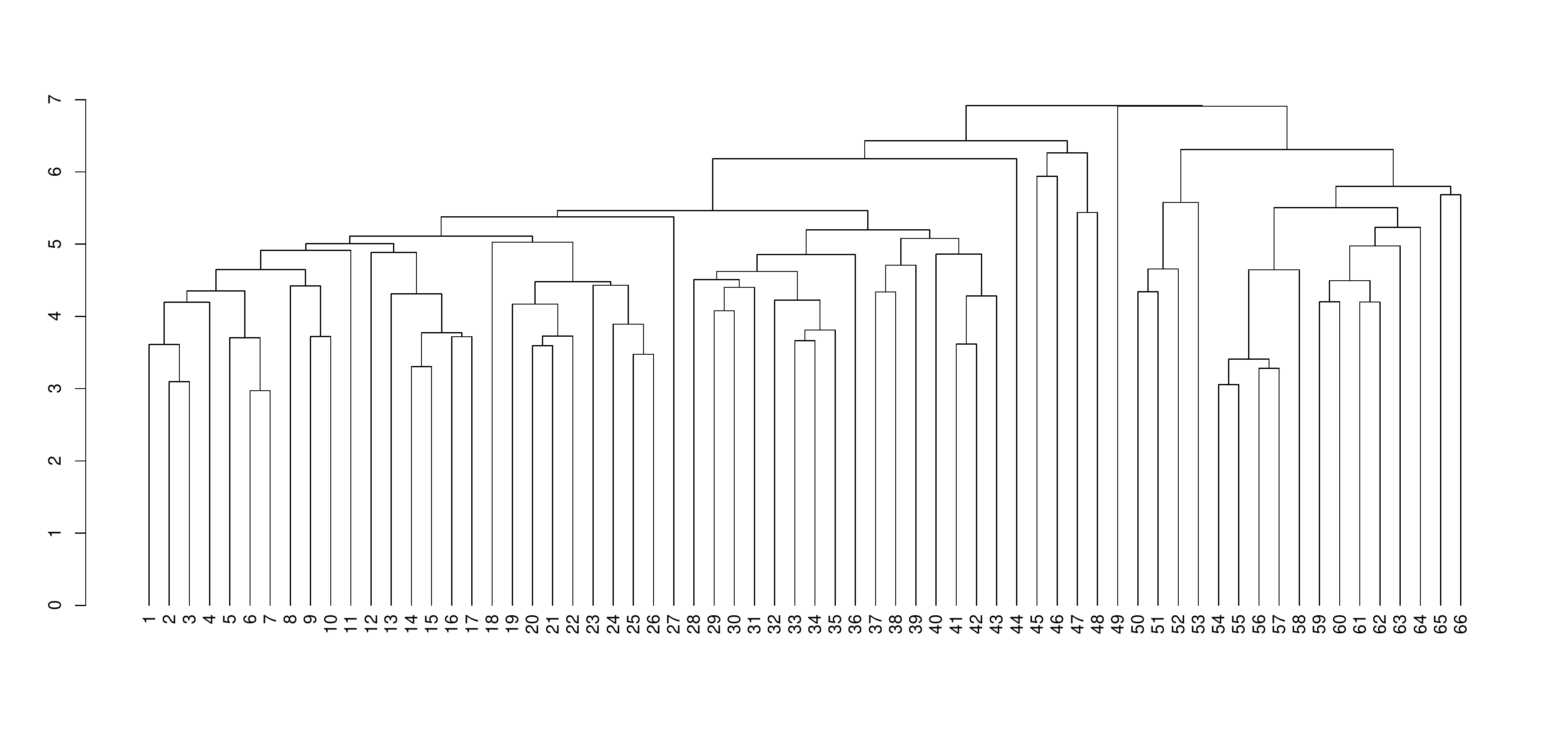}
\caption{Structure of {\em The Shadow Hours}, second snapshot from drafts.}
\label{mden3}
\end{figure}

Examining Figure~\ref{mden3}, which shows a slightly later draft of the same novel, in this draft section 44 is still a clearly anomolous section, but by much less of a degree, and 26, 27, and 6, now merge much more closely with the surrounding chapters. This is compared with contemporaneous notes from the project showing that 44 had now been drafted, and the other scenes were going though second drafts.  In this case the dendrogram allowed an ``at a glance'' notification of areas that required particular attention and revealed that a section had been missed due to a communication error in the team. 

\subsubsection{The Writer's Desk (TWD)} 
Given the much greater amount of time that staff at TWD had to examine a manuscript, the ability to ``immediately evaluate'' the structure of a document was less important. Instead the structural diagrams were used to validate, and later guide, the reviewer's own evaluation.   During the early stage of the project, staff reviewed documents as normal, and then examined the structural diagrams to see how much their interpretation of the diagram agreed with their  interpretation of the text.  As trust built, this progressed to reviewing documents before using the diagrams to check that no obviously anomoulous sections were missing, and then to reviewing both the text and the diagram at the same time, allowing the reviewer to re-examine text on the fly and get a much stronger impression of not only where the current section of text is going but how it slots into the overall narrative.

\section{Discussion and Future Work} 
\label{conclusions}
We have developed tools that we have used effectively to augment and improve upon qualitative analyses of narrative. Our  findings are that these techniques can be effective, depending greatly on the situation they are applied in.  Given the reported benefits of data visualisation~\cite{tufte1983visual,spiegelhalter2011visualizing}, the publishing sector has been slow to engage in use of visualisation.  In a set of interviews with 14 industry representatives that were conducted as part of the research, without exception the interviewees reported no use of software for anything other than counting words, and only a fraction of the interviewees were interested in seeing demos of any kind of supporting technology.   However, some publishing staff have been very positive about the idea of at-a-glance market placement and the added-value of being able to check that the section of the book that one has read is typical of the author's voice. Those who have made use of the technology are positive, and provided us with 
testimonials.  

\section{Acknowledgements}
We would like to thank especially Adam Ganz for his guiding expertise and long association along with  the staff at TWD, in particular Jacqueline Kibby. Thanks are also due to all participants on the TooManyCooks projects, and to the varied expertise provided by David Wells, Tony Greenwood, Adam Roberts, Meg Mitchell, Lucy Yeomans, Emm Johnstone, Patrick Leman, Yvonne Skipper, Peter Dunsmuir, John Vines, and Mark Dorling.

\bibliographystyle{plain}

\end{document}